\documentclass[aps,prb,twocolumn]{revtex4}

\usepackage{amssymb}
\usepackage{graphics}
\usepackage{graphicx}
\usepackage{amsmath}

\usepackage{times} 

\begin{document}

\title{Supercritical bifurcation of a hula hoop}

\author{Fr\'ed\'eric Moisy}
\email{moisy@fast.u-psud.fr}
\affiliation{FAST, B\^at.  502, Campus Universitaire, 91405 Orsay
Cedex, France}

\date{\today }

\begin{abstract}
The motion of a hoop hung on a spinning wire provides an illustrative
and pedagogical example of a supercritical bifurcation.  Above a
certain angular velocity threshold $\Omega_{c}$, the hoop rises,
making an angle $\theta \simeq (\Omega-\Omega_{c})^{1/2}$ with the
vertical.  The equation of motion is derived in the limit of a long
massless wire, and the calculated steady states are compared to
experimental measurements.  This simple experiment is suitable for
classroom demonstration, and provides an interesting alternative to
the classical experiment of the bead sliding on a rotation hoop.

\end{abstract}

\maketitle

\section{Introduction}

The rotation of rigid bodies often displays interesting instability
problems.  Bodies with three different inertia moments are well known
to have unstable rotation about the intermediate
axis~\cite{Greenwood}, as commonly observed from acrobatic jumps or
dives.  Throwing a tennis racket provides an easy illustration of this
instability of the free rotation.  Such purely inertial instability
has no threshold, {\it i.e.} it can be observed for very low rotation
rates.  In contrast, for axisymmetric bodies such as hoops, disks,
rockets, eggs etc., the free rotations remain stable, but different
mechanisms may also lead to instability when external forces are
present.  Under some circumstances, spectacular and unexpected
instabilities may originate from frictional forces.  This is the case
for the tippe-top, a popular toy that flips over and rotates on its
stem, or for the hard-boiled egg problem, which has recently received
a nice analysis~\cite{Moffatt02}.  More classically, the competition
between gravity and centrifugal forces may lead to an instability with
a finite rotation rate threshold~\cite{Drazin}, as illustrated by the
simple experiment described in this paper.

A hoop is hung on a long wire, whose upper end is spun.  At low
rotation rate, the hoop is vertical and simply spins about its
diameter.  Increasing the rotation rate, the hoop progressively rises
and becomes horizontal, spinning about its symmetry axis.  This
situation may appear paradoxical, since the horizontal position
maximizes both kinetic and potential energy.  It is similar to the
conical pendulum problem~\cite{Greenwood}, commonly illustrated in the
classical demonstration experiment of the bead sliding along a
vertically rotating hoop~\cite{Siva,Fletcher,Mancuso}.  The popular
`hula hoop' game, where the wire rotation is replaced by the hips
oscillations of the player, is a common illustration of this
phenomenon.  Once the hoop spins horizontally, its rotation is
maintained by a parametric oscillation mechanism~\cite{Caughey}. 

In this paper, the equation of motion is derived from the Lagrange's
equation.  An alternate derivation, from the angular momentum
equation, is also presented as a good illustration of dynamics of a
rigid body.  The hoop is shown to rise following a supercritical
bifurcation for the wire rotation rate $\Omega$ above a critical value
$\Omega_{c}=(2g/R)^{1/2}$, where $g$ is the gravitational acceleration
and $R$ the hoop radius.  The stable solution $\theta \sim
(\Omega-\Omega_{c})^{1/2}$ coincides with that of the conical pendulum
or the bead-on-a-hoop.  The nonlinear oscillations are briefly
described by means of phase portraits, and compared to that of the
bead-on-a-hoop.  Simple considerations from wire torsion and air
friction allow us to estimate the startup and damping timescales. 
Finally some experimental measurements are reported, and are shown to
compare well with the exact solution.

\begin{figure}
\centerline{
\includegraphics[width=3.3cm]{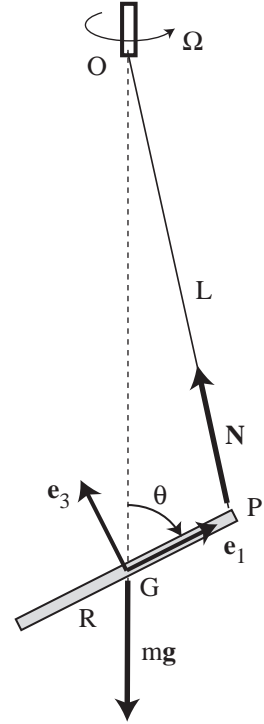}}
\caption[]{A hoop of center of mass G, attached to a wire OP, spins
about the vertical axis OG, making a pitch angle $\theta$ with the
vertical.}
\label{fig:sketch}
\end{figure}

In this experiment, the angular rotation threshold being of order of a
few rad/s, the wire can be simply spun with the fingers.  Since it
only requires a wire and a hoop, this experiment can be conveniently
used as a classroom illustration of spontaneous symmetry breaking and
bifurcation.  Although the detail of the calculation is somewhat more
subtle than that of the classical bead-on-a-hoop problem, since it
deals with rigid body dynamics, the physics is basically the same and
does not require the much heavier apparatus of the bead-on-a-hoop
demonstration.

\section{Theory}

\subsection{Equation of motion}

We consider a thin uniform hoop, of radius $R$ and mass $m$, fixed at
a point P of its periphery by a massless wire of length $L \gg R$ (see
figure~\ref{fig:sketch}).  The wire is spun from its other end O at a
constant angular velocity $\Omega$.  The center of mass G is assumed
to remain at the vertical of O. Let us first consider a torsionless
wire, so that the hoop angular velocity $\dot{\phi}$ about the
vertical axis ${\bf e}_{z}$ instantaneously follows the imposed
angular velocity $\Omega$.

Let us consider the rotating frame of reference of the hoop $({\bf
e}_{1},{\bf e}_{2},{\bf e}_{3})$.  We note $\theta$ the pitch angle
between the hoop plane and the vertical axis ${\bf e}_{z}$.  We
consider for simplicity a zero roll angle about ${\bf e}_{1}$ ({\it
i.e.} we assume that the point P remains the highest point of the
hoop), so that $\theta$ is the only degree of freedom for this
problem.  The absolute angular velocity {\boldmath $\omega$} of the
hoop has two contributions.  The first one, of magnitude $\Omega$, is
imposed from O through the torsionless wire, and is about ${\bf
e}_{z}=\cos \theta \, {\bf e}_{1} + \sin \theta \, {\bf e}_{3}$ (the
points O and G are at rest, so that OG is the instantaneous axis of
rotation when $\theta$ is kept constant).  The second one comes from
the pitch variation $\dot{\theta}$, and is about the axis ${\bf
e}_{2}$.  We therefore obtain
\begin{equation}
\mbox{\boldmath $\omega$} = \left( \begin{array}{l} \Omega \cos \theta \\
\dot{\theta} \\ \Omega \sin \theta \end{array} \right).
\label{eq:aav}
\end{equation}

The potential energy is $V = m g z_{G}$, where $z_{G} \simeq R (1-\cos
\theta)$ is the center of mass elevation to first order in $R/L$.  The
kinetic energy $T$ has two contributions: one from the rotation
$\frac{1}{2} \mbox{\boldmath $\omega$} \cdot \tilde{I} \cdot
\mbox{\boldmath $\omega$}$, where $\tilde{I}$ is the inertia matrix of
the hoop relative to G, and one from the vertical translation of the
center of mass, $\frac{1}{2} m \dot{z}_{G}^{2} \simeq \frac{1}{2} m
R^{2} \dot{\theta}^{2} \sin^{2} \theta$.  In the reference frame of
the hoop, the inertia matrix $\tilde{I}$ has diagonal components
$I_{1} =I_{2} = \frac{1}{2} mR^{2}$ and $I_{3} = mR^{2}$.  The
Lagrangian function ${\cal L} = T - V$ then finally writes
\begin{equation}
{\cal L} = \frac{1}{4} m R^{2} \left[ \Omega^{2} (1 + \sin^2
\theta)  + \dot{\theta}^{2} (1+2 \sin^{2} \theta) \right] 
+ m g R \cos \theta.
\end{equation}
Writing the Lagrange's equation for the coordinate $\theta$, 
we end up with the differential equation of motion
\begin{equation}
\ddot{\theta} (1+2 \sin^{2} \theta) - (\Omega^{2} - 2
\dot{\theta}^{2}) \sin \theta \cos \theta + \Omega_{c}^{2} \sin \theta
= 0.
\label{eq:edc}
\end{equation}
where $\Omega_{c} = \sqrt{2 g /R}$. 

This equation of motion can also be obtained from the angular momentum
equation and Newton's law.  In the rotating frame of reference, the
angular momentum equation reads
\begin{equation}
\frac{d {\bf L}} {dt} + \mbox{\boldmath $\omega$} {\times} {\bf L} =
{\bf \Gamma},
\label{eq:eulv}
\end{equation}
where ${\bf L} = \tilde{I} \mbox{\boldmath $\omega$}$ is the angular
momentum and ${\bf \Gamma} = {\bf GP} {\times} {\bf N}$ is the torque
of the wire tension relative to the center of mass (the gravitational
torque vanishes).  The wire tension is obtained from Newton's law,
\begin{equation}
m \ddot{z}_{G} \, {\bf e}_{z} = {\bf N} + m {\bf g}.
\end{equation}
To first order in $R/L$, ${\bf N}$ is vertical, so its torque is
simply ${\bf \Gamma} \simeq - RN \sin \theta \, {\bf e}_{2}$.  Using
$z_{G} \simeq R(1-\cos \theta)$, we obtain
\begin{equation}
{\bf \Gamma} \simeq -R \left( mg+mR(\dot{\theta}^{2} \cos \theta +
\ddot{\theta} \sin \theta) \right) \sin \theta \, {\bf e}_{2}.
\end{equation}
Replacing into (\ref{eq:eulv}) and projecting on ${\bf e}_{2}$, we
finally recover the equation of motion~(\ref{eq:edc}).

\subsection{Steady states and small oscillations}

We are interested in the steady states and the natural frequency
for small oscillations about them.  To first order in $\theta$,
equation~(\ref{eq:edc}) reduces to
\begin{equation}
\ddot{\theta} - \Omega^{2} \sin \theta \cos \theta + \Omega_{c}^{2}
\sin \theta = 0.
\label{eq:edl}
\end{equation}
It is worth noting that this linearized equation is exactly the same
as the one from the bead-in-a-hoop problem~\cite{Siva}.  The
difference between equations~(\ref{eq:edc}) and~(\ref{eq:edl})
originates from the kinetic energy of the center of mass translation,
which is not present in the bead-on-a-hoop.

In addition to the trivial solution $\theta_{\mathrm{eq}}=0$,
equation~(\ref{eq:edl}) has a non trivial solution for $\Omega \geq
\Omega_{c}$,
\begin{equation}
\theta_{\mathrm{eq}} = \pm \cos^{-1} \left( \frac{\Omega_{c}}{\Omega}
\right)^{2}.
\label{eq:sol}
\end{equation}
Following the usual terminology, the pitch angle
$\theta_{\mathrm{eq}}$ is the order parameter, and we introduce the
reduced angular velocity $\epsilon = \Omega/\Omega_{c} - 1$ as the
control parameter.  The linearization of~(\ref{eq:sol}) finally leads
to the classical form of a supercritical pitchfork bifurcation for
$\epsilon \ll 1$:
\begin{equation}
\theta_{\mathrm{eq}} \simeq 2 \sqrt{\epsilon}.
\label{eq:solap}
\end{equation}

Stability and natural frequency for small oscillations are obtained by
introducing in equation~(\ref{eq:edl}) a small perturbation in the
form
\begin{equation}
\theta(t) = \theta_{\mathrm{eq}} + \delta \theta \, e^{(\sigma + i
\omega) t},
\end{equation}
where $\theta_{\mathrm{eq}}$ stands for the trivial or non-trivial
solution.  The trivial solution is stable for $\Omega < \Omega_{c}$
($\sigma=0$), with a natural frequency $\omega = (\Omega_{c}^2 -
\Omega^{2})^{1/2}$, and unstable for $\Omega > \Omega_{c}$, with a
growth rate given by $\sigma = (\Omega^2 - \Omega_{c}^{2})^{1/2}$ (and
$\omega=0$).  The non-trivial solution~(\ref{eq:sol}) is found to be
always stable, with the same natural frequency.  In terms of
$\epsilon$, one can see that the period of the oscillations,
\begin{equation}
T \sim |\epsilon|^{-1/2},
\end{equation}
diverges as one approaches the transition from both side.  This
critical slowing down is a usual signature of supercritical
bifurcation\cite{Drazin}.


\subsection{Nonlinear oscillations}

\begin{figure}
\centerline{
\includegraphics[width=7.5cm]{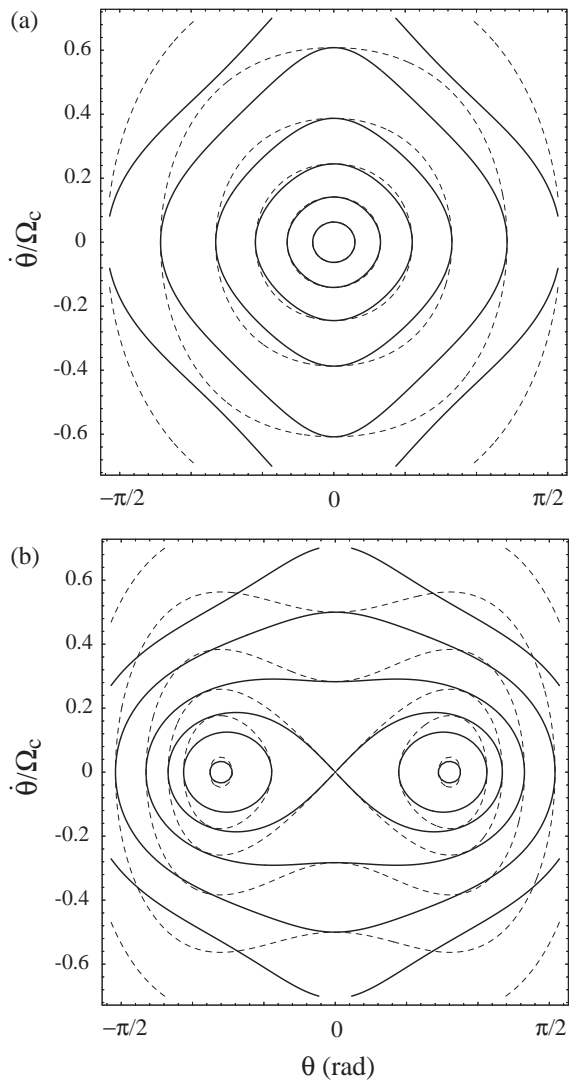}}
\caption[]{Orbits in the phase space $(\theta,
\dot{\theta}/\Omega_{c})$ for the hula hoop (solid lines) compared to
the bead-on-a-hoop system (dashed lines).  ($a$): $\Omega=0.8
\Omega_{c}$.  ($b$): $\Omega=1.2 \Omega_{c}$.}
\label{fig:orbits}
\end{figure}

When the oscillation amplitudes about the steady states are not small,
the nonlinear terms in~(\ref{eq:edc}) become important and may affect
the dynamics of the system.  The orbits in the phase space $(\theta,
\dot{\theta})$ then provide a useful tool to characterize the
nonlinear dynamics of the hula hoop, and to compare it with the one of
the bead-on-a-hoop~(\ref{eq:edl}).  The equation of motion can be
integrated using the Painlev\'e invariant,
\begin{equation}
C = \dot{\theta} \frac{\partial \cal L} {\partial \dot{\theta}} - 
{\cal L},
\end{equation}
which takes a constant value along an orbit (this invariant
corresponds to the total energy $E=T+V$ in the case of a conservative
system).  Figure~\ref{fig:orbits} shows isolines of $C$ for two
forcing frequencies, below ($\Omega/\Omega_{c}=0.8$, fig.~$a$) and
above ($\Omega/\Omega_{c}=1.2$, fig.~$b$) the transition, together
with the orbits of the bead-on-a-hoop problem (dashed lines).  One can
clearly see from these phase portraits the trivial solution
$\theta_{\mathrm{eq}}=0$ for $\Omega/\Omega_{c}=0.8$ and the two non
trivial solutions~(\ref{eq:sol}) separated by a saddle point at
$\theta=0$ for $\Omega/\Omega_{c}=1.2$.  Note that we are only
interested in the closed orbits in the domain $|\theta| \leq \pi/2$:
orbits that cross $\theta = \pm \pi/2$ are not consistent with the
assumption that the point P remains the highest point of the hoop.

Around the stable fixed points, both the two systems show nearly
elliptic orbits, as expected from small harmonic oscillations.  Larger
oscillations of the hula hoop show orbits with sharper corners around
$\theta \simeq 0$, associated to larger angular velocities
$|\dot{\theta}|$.  This discrepancy originates from the vertical
translation of the center of mass, which is responsible for an
additional pitch angle acceleration when the hoop is nearly vertical.

\subsection{Timescales}

Two timescales are relevant for a practical experiment: the startup
timescale of rotation $\tau_{s}$ and the damping timescale of
oscillations $\tau_{d}$.  The startup timescale may be obtained
considering the wire torsion.  We start from an initially vertical
hoop ($\theta=0$), and let now its rotation angle $\phi$ about ${\bf
e}_{z}$ be free.  The wire communicates a torque $\kappa (\phi-\Omega
t) = \kappa \, \delta \phi$ to the hoop, where $\kappa$ is the torsion
constant and $\delta \phi$ the wire torsion.  The angular momentum
equation about ${\bf e}_{z}$ applied to the hoop then reads
\begin{equation}
\frac{1}{2} m R^{2} \delta \ddot{\phi} +\kappa \, \delta \phi =0,
\end{equation}
leading to a startup timescale $\tau_{s} \simeq R (m/2\kappa)^{1/2}$
(which corresponds also to the period of the wire torsion oscillations
if no damping were present).

The damping timescale $\tau_{d}$ is relevant both for the $\phi$
oscillations due to the wire torsion, and for the $\theta$
oscillations around the equilibrium states.  It can be obtained
considering the dissipation with the surrounding air.  With typical
velocity of order m/s and hoop thickness $w$ of a few millimeters, we
can make use of a turbulent estimate for the drag force.  Neglecting
$\dot{\theta}$ compared to $\dot{\phi}$, a unit length $dl$ of the
hoop experiences a drag force $df \simeq \rho R^{2} \dot{\phi}^{2} w
\, dl$, where $\rho$ is the air density.  Integrating over the hoop
perimeter leads to the frictional torque
\begin{equation}
{\bf \Gamma}_{r} \simeq - \rho R^{4} \dot{\phi}^{2} w \, \frac{{\bf
\Omega}}  {|{\bf \Omega}|}.
\end{equation}
The angular momentum equation about ${\bf e}_{z}$ reads
\begin{equation}
\ddot{\phi} + \left( \frac{\rho_{0}}{\rho} \frac{w}{R} \right)
\dot{\phi}^{2} =0.
\end{equation}
where $\rho_{0}$ is the hoop density.  Taking $\dot{\phi} \simeq
\Omega$ and $\ddot{\phi} \simeq \Omega / \tau_{d}$, the damping
timescale finally writes
\begin{equation}
\tau_{d} \simeq \left( \frac{\rho_{0}}{\rho} \frac{w}{R} \right)
\Omega^{-1}.
\end{equation}
For values of practical interest, $\tau_{d}$ is of order of a few tens
of the rotation period $\Omega^{-1}$.  Note here that in the case of
the real hula hoop game, the rolling friction on the player hips is
fortunately dominant, leading to a damping timescale of order
$\Omega^{-1}$.

\section{Experimental results}

\begin{figure}[hbtp]
\centerline{
\includegraphics[width=7.8cm]{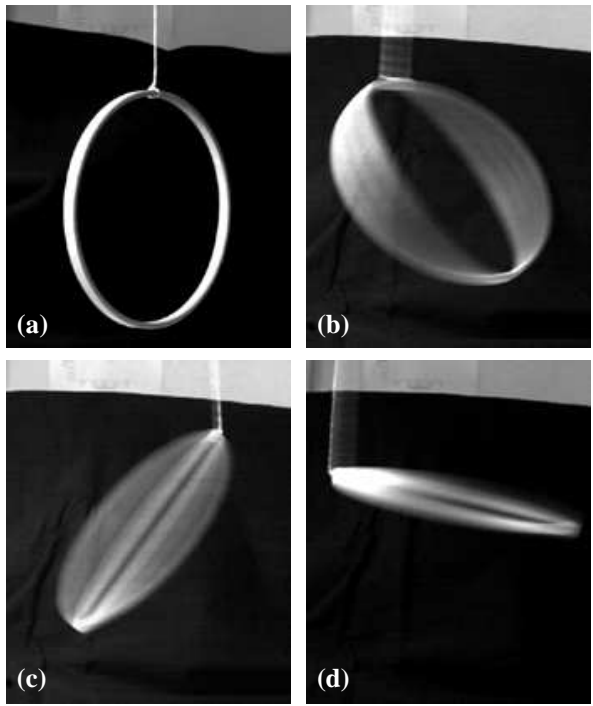}}
\caption[]{Pictures of the hoop.  (a)~The hoop at rest.  (b)~and
(c)~$\theta \simeq 37^\mathrm{o}$.  (d)~$\theta \simeq
78^\mathrm{o}$.}
\label{fig:pic_hula}
\end{figure}

An experiment has been carried out using a wood hoop ($\rho_{0} \simeq
0.67$~g.cm$^{-3}$) of mass $m=15.5$~g, radius $R=86$~mm and section $4
\times 10$~mm$^{2}$.  The expected frequency threshold is then
$\Omega_{c}/2 \pi \simeq 2.40$~Hz (classical hula hoops have
$\Omega_{c}/2 \pi \simeq 1$~Hz).  A simple cotton thread, of length
$L=0.68$~m and mass less than 3\% of the hoop mass, was fixed to a
constant current motor.  The startup timescale for this wire can be
estimated from the free oscillation period, $\tau_{s} \simeq 5$~s
(corresponding to a torsion constant $\kappa \simeq 2 \times
10^{-6}$~N.m).  This wire is far from being torsionless, and during
the early stage of the rotation, torsional energy is stored into the
wire.  The wire progressively communicates rotation to the hoop, and
the fluctations (due to a varying imposed rotation rate at its upper
end, or to pitch angle variation at the other end) are smoothed down
on a timescale $\tau_{d}$.


Seen from the side, the hoop rotation appears as Lissajous ellipses,
whose principal axis makes an angle $\theta$ with the vertical axis. 
Pictures acquired from a simple CCD camera (see
figure~\ref{fig:pic_hula}) allows us to easily measure the pitch angle
$\theta$ within 1$^\mathrm{o}$ from these ellipses.  The time aperture
of the camera may blur the picture, but facilitate in some case the
angle measurement (see for instance picture c).

\begin{figure}[tbh]
\vskip 2mm
\centerline{
\includegraphics[width=8.4cm]{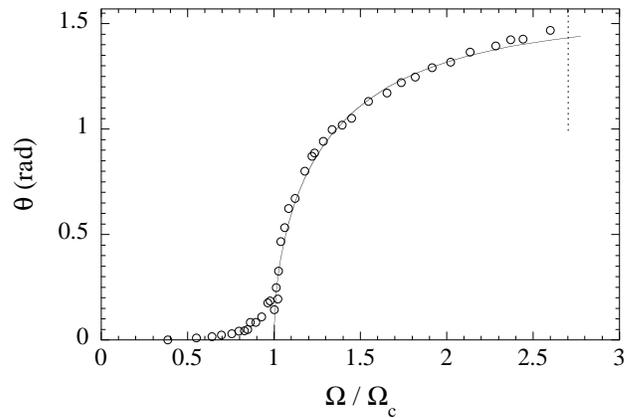}}
\caption[]{Bifurcation diagram of the pitch angle $\theta$ as a
function of the normalized angular velocity $\Omega/\Omega_{c}$.  The
solid line is the calculated solution~(\ref{eq:sol}), with
$\Omega_{c}$ adjusted to fit the experimental data.}
\label{fig:avsw}
\end{figure}

The measured pitch angles $\theta$, shown in figure~\ref{fig:avsw},
are in excellent agreement with the exact solution~(\ref{eq:sol}),
except for low rotation rate, where non zero angles are measured below
the expected transition.  The experimental frequency threshold,
obtained by extrapolating the curve down to $\theta \simeq 0$, is
$\Omega_{c}/2 \pi \simeq (2.30 \pm 0.03)$~Hz, which agrees within 5\%
with the theoretical value.  The discrepancy at low rotation rate is
typical of an imperfect bifurcation, where small asymmetry in the
apparatus (such as the position of the knot) slightly anticipates the
destabilization of the basic state.

An interesting observation is that, for $\Omega \geq 2.6 \Omega_{c}$
(see the dashed line in figure~\ref{fig:avsw}), the bifurcated state
is no stable any more: a secondary instability appears, in the form of
a slow precession of the center of mass, with a period of about ten
rotation periods.  This new behavior is probably an effect of the non
zero mass of the wire, and can clearly not be described in our
calculation, where the hoop center of mass is constrained to remain
aligned with the vertical axis.

\begin{figure}[t]
\centerline{
\includegraphics[width=7.5cm]{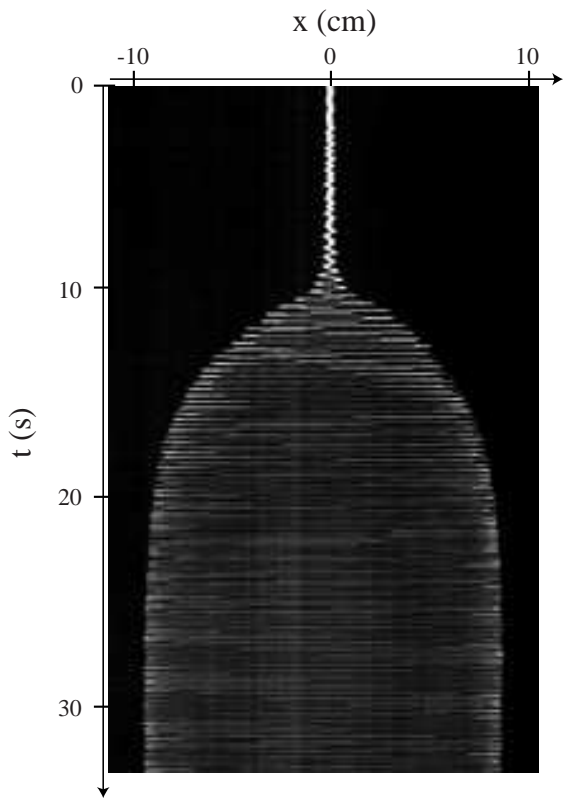}}
\caption[]{Spatio-temporal diagram of the light intensity recorded on
an horizontal line, showing the wire oscillation, when the frequency
$\Omega/2\pi$ is increased from 0 to 5.5~Hz.}
\label{fig:spatio}
\end{figure}

Transient phenomena may also be investigated, by means of usual video
processing.  An illustration is given in figure~\ref{fig:spatio},
showing a spatio-temporal diagram obtained by collecting the light
intensity recorded on a horizontal line passing through the wire.  In
this example, the imposed frequency $\Omega/2\pi$ has been suddenly
increased from 0 to 5.5~Hz~$\simeq 2.4~\Omega_{c}/2\pi$ (first arrow). 
One can see that, after a transient time of around 7~s (second arrow)
during which the wire rotation propagates down to the hoop, the
amplitude increases and saturates to a finite value.

\begin{figure}[t]
\vskip 2mm
\centerline{
\includegraphics[width=8cm]{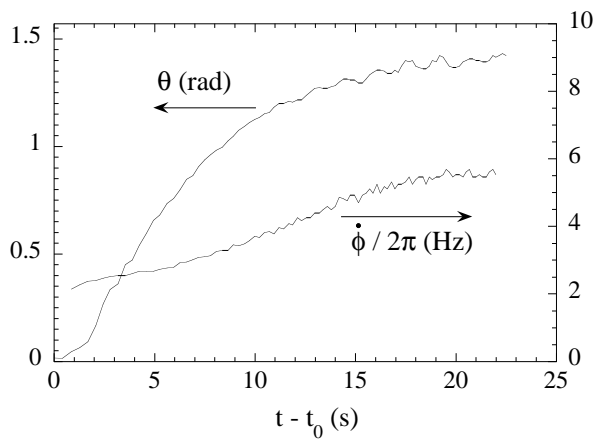}}
\caption[]{Instantaneous pitch angle $\theta$ (left) and rotation
frequency $\dot{\phi}/2\pi$ (right) as measured from
figure~\ref{fig:spatio}.  The origin time $t_{0}$ corresponds to the
second arrow in that figure.}
\label{fig:envel}
\end{figure}

From this diagram, the instantaneous pitch angle as well as the
instantaneous oscillation frequency $\dot{\phi}/2\pi$ may be
extracted, as shown in figure~\ref{fig:envel}.  The frequency is
obtained from averaging over 6 successive oscillations.  Due to the
progressive torsion of the wire, this frequency slowly approaches its
imposed value 5.5~Hz.  As a consequence, the increase of the pitch
angle towards its stationary value $\theta_{\mathrm{eq}} \simeq
1.42$~rad is rather slow.

\begin{figure}[t]
\vskip 2mm
\centerline{
\includegraphics[width=8.4cm]{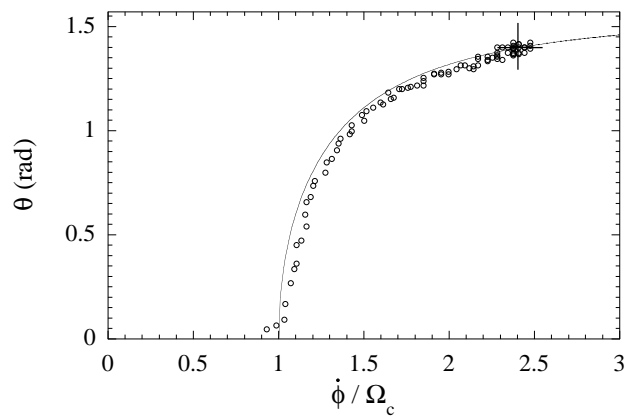}}
\caption[]{Instantaneous pitch angle $\theta$ as a function of the
instantaneous rotation frequency $\dot{\phi}/\Omega_{c}$ (same data as
in figure~\ref{fig:envel}).  The solid line is the calculated
solution~(\ref{eq:sol}), as in figure~\ref{fig:avsw}.  The cross
indicates the stationary solution $\theta_{\mathrm{eq}}\simeq
1.42$~rad for $\Omega=2.4~\Omega_{c}$.}
\label{fig:bifinst}
\end{figure}

It is interesting to note that, when plotting the instantaneous pitch
angle as a function of the instantaneous frequency (see
figure~\ref{fig:bifinst}), the bifurcation diagram of
figure~\ref{fig:avsw} is recovered to a high degree of accuracy.  This
suggests that the pitch angle follows `adiabatically' the
instantaneous frequency, so that the apparent growth rate is
essentially controlled by the wire torsion more than the intrinsic
dynamics of the instability (at least far from the transition).

\section{Discussion}

This simple experiment provides an interesting alternative to the
bead-on-a-hoop~\cite{Siva,Fletcher,Mancuso} experiment as a mechanical
analog to bifurcation and second-order phase transition in physics. 
The poor attention paid to the choice of the material and the
experimental conditions illustrates the robustness of the phenomenon,
and makes this experiment easy to work out by undergraduate students. 
Further experiments can be performed, {\it e.g.} studying the
transient phenomena resulting from a small perturbation.  As a
suggestion, restrain the point P on the vertical axis when $\Omega >
\Omega_{c}$ by means of a small hook around the wire.  Releasing the
hook allows us to measure the growth rate and to characterize its
divergence as $\epsilon \rightarrow 0$.  This method can be hardly
achieved with a bead-on-a-hoop apparatus due to its inherent difficult
access.

Another motion may compete with the bifurcation described in this
paper: the hoop can rotate as a whole about the vertical axis, its
center of mass remaining aligned with the wire.  This is the usual
motion for the conical pendulum, which also leads to a supercritical
bifurcation with a threshold simply given by the natural frequency
$\omega_{0} \simeq \sqrt{g/(L+R)}$.  The conical pendulum motion
overcomes the hula hoop motion for $\omega_{0}$ of order of
$\Omega_{c}$, {\it i.e.} for a wire length of order of the hoop
radius.  Moving away the center of mass from the vertical axis may
allow us to observe the competition between the two regimes, even for
a longer wire.

Similar bifurcations as the result of a competition between
centrifugal force and gravity are present in a number of situations. 
Spinning plates provides an interesting illustration: as for the real
hula hoop, the motion here is forced by the precession of the rigid
rod rather than its rotation.  The imposed precession frequency has to
overcome the natural frequency for the plate to stand up, and then the
horizontal state is maintained by a parametric oscillation
mechanism~\cite{Caughey}.  Lasso roping is another example, where the
rigid hoop is replaced by a deformable loop.  Here again the motion is
maintained by the precession of the spoke rather than rotation.  As a
consequence, the knot on the loop makes the spoke to rotate as well,
so that the cord has to be continuously untwisted at its other end.

It is worth pointing out that this experiment can be carried out with
any rigid body, not necessarily axisymmetric.  In this case, the angle
of the equilibrium state~(\ref{eq:sol}) remains unchanged, and the
angular velocity threshold just becomes $\Omega_{c} = \sqrt{mga /
(I_{3}-I_{1})}$, where $a=|{\bf GP}|$ is the distance between the
center of mass and the knot.  Note however that, for bodies with three
different inertia moments, in addition to the bifurcation described
here, inertial instabilities may also occur.  Such instability
involves the roll angle about ${\bf e}_{1}$ as an additional degree of
freedom, and is not described by our calculation.

\acknowledgments This work has benefited from fruitful discussions
with M.~Rabaud and B.~Perrin.

\end{document}